\documentclass[conference]{IEEEtran}

\IEEEoverridecommandlockouts
\usepackage{cite}
\usepackage{amsmath,amssymb,amsfonts}
\usepackage{algorithmic}
\usepackage{graphicx}
\usepackage{textcomp}
\usepackage{xcolor}
\def\BibTeX{{\rm B\kern-.05em{\sc i\kern-.025em b}\kern-.08em
    T\kern-.1667em\lower.7ex\hbox{E}\kern-.125emX}}

\usepackage{graphicx}
\usepackage{multirow}
\usepackage{float}
\usepackage{subcaption} 
\usepackage[linesnumbered,ruled,vlined]{algorithm2e}
\usepackage{framed, blindtext}
\usepackage{tablefootnote}
\usepackage{tikz}
\usepackage{booktabs}
\usepackage[utf8]{inputenc}
\usepackage{amsmath}
\usepackage{amssymb}
\usepackage{amsfonts}
\usepackage{url}

\usepackage[framemethod=tikz]{mdframed}

\usepackage{booktabs}

\mdfdefinestyle{mpdframe}{
    frametitlebackgroundcolor   =black!15,
    frametitlerule              =true,
    roundcorner                 =5pt,
    middlelinewidth             =1pt,
    innermargin                 =0.1cm,
    outermargin                 =0.1cm,
    innerleftmargin             =0.1cm,
    innerrightmargin            =0.1cm,
    innertopmargin              =0.1cm,
    innerbottommargin           =0.1cm
}

\begin{document}

\title{XMD: An expansive Hardware-telemetry based Mobile Malware Detector for Endpoint Detection}

\author{
{\rm Harshit Kumar},
{\rm Biswadeep Chakraborty},
{\rm Sudarshan Sharma},
{\rm Saibal Mukhopadhyay}\\
Dept. of Electrical and Computer Engineering\\
Georgia Institute of Technology, {Atlanta, USA}\\
\rm \{hkumar64, biswadeep, ssharma497\}@gatech.edu, saibal@ece.gatech.edu\\
} % end author
% \author{\IEEEauthorblockN{Harshit Kumar, Biswadeep Chakraborty, Sudarshan Sharma, Saibal Mukhopadhyay}
% \IEEEauthorblockA{\textit{Dept. of Electrical and Computer Engineering \\ Georgia Institute of Technology, Atlanta, Georgia, USA}}}
% \author{\IEEEauthorblockN{Anonymous Authors}}

\maketitle
\begin{abstract}
Hardware-based Malware Detectors (HMDs) have shown promise in detecting malicious workloads. However, the current HMDs focus solely on the CPU core of a System-on-Chip (SoC) and, therefore, do not exploit the full potential of the hardware telemetry. In this paper, we propose XMD, an HMD that uses an expansive set of telemetry channels extracted from the different subsystems of SoC. XMD exploits the thread-level profiling power of the CPU-core telemetry, and the global profiling power of non-core telemetry channels, to achieve significantly better detection performance than currently used Hardware Performance Counter (HPC) based detectors. We leverage the concept of manifold hypothesis to analytically prove that adding non-core telemetry channels improves the separability of the benign and malware classes, resulting in performance gains. We train and evaluate XMD using hardware telemetries collected from 723 benign applications and 1033 malware samples on a commodity Android Operating System (OS)-based mobile device. XMD improves over currently used HPC-based detectors by 32.91\% for the in-distribution test data. XMD achieves the best detection performance of 86.54\% with a false positive rate of 2.9\%, compared to the detection rate of 80\%, offered by the best performing signature-based Anti-Virus(AV) on VirusTotal, on the same set of malware samples. 
\end{abstract}


\begin{figure*}[!h]
% \vspace{-5mm}
\captionsetup{justification=centering}
    \centering
    \includegraphics[scale = 0.84]{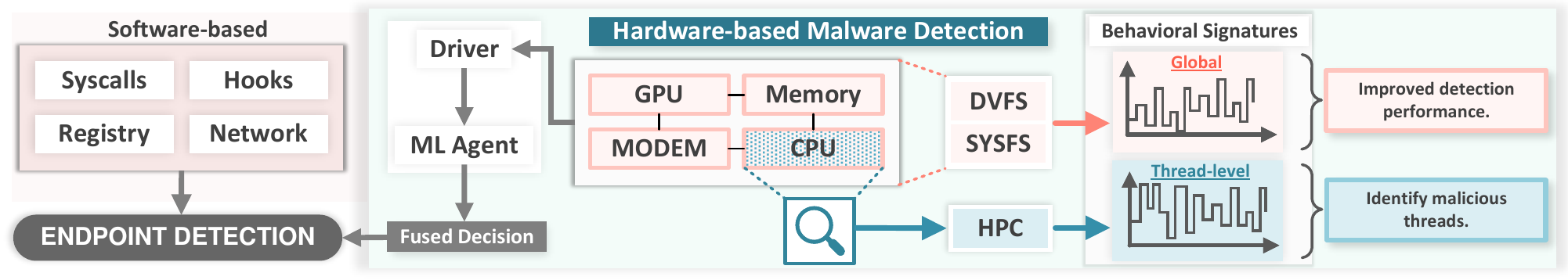}
    \caption{Comprehensive cross-stack monitoring performed by modern Endpoint-Detection : [Left] Software-based malware detection approaches, [Right] Proposed XMD operating on expansive set of telemetry channels.}
    % \vspace{-6mm}
    \label{fig:hmd_endpoint_big_picture}
\end{figure*}

\section{Introduction}
The previous decade has witnessed an explosive growth of malicious applications, compromising the security of modern devices \cite{fireeye}. Consequently, Endpoint Security has moved towards behavior analysis that involves continuous monitoring of sensors across the compute stack and rigorous data analysis \cite{microsoft_endpoint,8681127}. As shown in Figure \ref{fig:hmd_endpoint_big_picture}, behavior analysis techniques monitor the program's execution using semantically rich information sources like registry keys, network endpoints, system calls, and operating system (OS) hooks. However, malicious actors can potentially subvert such protection mechanisms by tampering with the software telemetry \cite{intel_tdt, Karantzas_2021}. Moreover, such software-level detection approaches result in significant performance overhead. Hence, there has been a recent thrust in data-driven approaches for detecting malicious workloads using low-level hardware telemetry, which promises low overheads and better resilience against tampering compared to software-based telemetry \cite{demme_hpc}.

The application of hardware-level telemetry, like HPC, energy telemetry channels (e.g., Intel’s RAPL), and Dynamic Voltage and Frequency Scaling (DVFS), towards malware detection, has recently gained interest \cite{kuruvila_efficiency_hmd,sanyal_hmd_reservoir,hmd_fine_grained,demme_hpc,analyze_hmd,quant_hmd,hpc_myth_or_fact,chawla_iotj,low_barrier_hmd,explainable_hmd,ensemblehmd,kumar_uncertainty,hpc_ransomware}. Monitoring the hardware telemetry provides visibility into active threats, even during the presence of anti-evasion techniques like obfuscation \cite{quant_hmd}, or cloaking in virtual machines \cite{intel_tdt}. As shown in Figure \ref{fig:hmd_endpoint_big_picture}, HMD, along with software-level detection techniques, is part of the commercially-deployed collaborative defense model, e.g., Endpoint Detection and Response (EDR), that detects and contains threats \cite{noauthor_qualcomm_2017,intel_tdt}. In such a collaborative system, HMDs can provide an additional layer of detection capability across the kill chains, complement other detectors, and restrict the ability of the adversary to move in the environment without triggering detection. 

Demme et al.  demonstrated the potential of using Hardware Performance Counters (HPCs) for detecting Android malware, Linux Rootkits, and side-channel attacks \cite{demme_hpc}. Thereafter, various methodologies were proposed for enhancing the predictive performance \cite{hmd_fine_grained, low_barrier_hmd,kuruvila_efficiency_hmd, ensemblehmd, kumar_uncertainty}, and to reduce the performance overhead of HMDs \cite{sanyal_hmd_reservoir,analyze_hmd}. Additionally, efforts have been made to delve into alternate CPU-telemetry sources \cite{chawla_iotj}. However, as highlighted by prior literature, one key drawback of using HPCs is the limited number of Hardware Counters, which consequently restricts the behavioral profiling power of such an approach. Botacin \& Gr\'egio recently introduced the concept of attack space that refers to the range of possible actions or techniques an attacker can utilize against a system \cite{hpc_malware_theory}. The authors argued that HPCs are more fitting for an attack space comprising of malicious workloads that exhibit architectural side-channel effects. We concur with their thesis and further assert that the exclusive reliance on CPU hardware telemetry only provides a partial view of the impact of running workloads on an SoC, thereby limiting the scope of attack space that an HPC-based HMD can cover. The CPU-telemetry-based approach does not leverage the vast potential of comprehensive hardware telemetry from \textit{non-core devices} of the SoC (e.g., Graphics Processing Unit (GPU), memory, buses, network). \textbf{In light of this, our work seeks to boost the detection performance of HPC-based HMDs when dealing with general-purpose malware, by leveraging hardware telemetry from non-core devices in the SoC.}

\textit{Proposed Work.} Our proposed approach, XMD, addresses the limitations in behavioral visibility inherent to HPC-based HMDs. These limitations arise primarily due to their exclusive reliance on CPU-based telemetry. By supplementing HPC-based telemetry with system-wide telemetry from non-core devices of the SoC, such as the GPU, memory, buses, and network, XMD notably improves upon the HPC-based HMDs. This system-wide telemetry data is extracted from DVFS signatures associated with non-core devices and low-level telemetry from select SYSFS nodes. Such an expanded telemetry set equips the ML agent with a broader understanding of a workload's impact on the entire SoC. This subsequently extends the attack space that the HMD can effectively cover, resulting in significant improvements in detection performance compared to prior HPC-based HMDs.

Our key innovation is grounded in the theorem developed by leveraging the concept of manifold hypothesis (Section \ref{sec:theory}). Prior works in deep learning have used manifold hypothesis for studying the geometry of manifolds in Deep Neural Networks (DNNs) for vision \cite{cohen2020separability}, audition \cite{stephenson2019untangling}, and language modeling \cite{mamou2020emergence,alleman2021syntactic}. Such analytical frameworks have explained the generalization performance of DNNs \cite{stephenson2021geometry}. In this work, we use manifold hypothesis to show that adding additional telemetry channels from non-core devices of the SoC increases the solution volume \cite{gardner1988space} of the ML classifier and leads to the superior classification performance of XMD. 

To empirically validate our approach, we created a bare-metal framework to automate the collection of hardware telemetry data using a commodity Android mobile device (Google Pixel-3). Using this data-collection framework, we developed the STD-Dataset which helped us empirically validate our primary hypothesis. Welch's t-test-based statistical analysis performed on the collected data revealed a significant increase in the distinguishability between benign and malicious classes as we incorporated hardware telemetry from various non-core devices. The increased distinguishability consequently led to an improvement in the predictive performance of ML classifiers. For instance, the F1-score achieved by using only CPU-based telemetry (i.e., HPC data), as in prior works, ranged from 0.66-0.76. However, by augmenting this CPU-based telemetry with our proposed non-core telemetry channels, we successfully elevated the predictive accuracy of the machine learning classifier to an F1-score of 0.90-0.92. These results underscore the utility of integrating telemetry from various non-core devices in improving the detection performance of an HMD.

To summarize, our main contributions are as follows:
\begin{itemize}
\item We develop a theorem using manifold hypothesis to prove that using telemetry channels from non-core devices improves the detection performance of the HMD.
\item We perform empirical experiments to validate the proposed theorem. We design a bare-metal data collection framework to collect expansive hardware telemetry that captures a workload's impact on different sub-systems of the SoC like CPU, GPU, Memory, Buses, and Network. Using the collected data, we show that using multiple telemetry channels from the different sub-systems of an SoC results in better classification performance when compared to classifiers that operate on the telemetry from a single sub-system, e.g., CPU-based telemetry. 
\end{itemize}

\section{Background and Threat Model}
\label{sec:background}
\subsection{Classes of Hardware Telemetry}
\textbf{Thread-level Profiling : Hardware Performance Counters (HPCs). }
HPCs are dedicated physical registers in modern processors that store the count of microarchitectural events in a CPU core during process execution. They were originally designed to identify performance bottlenecks. To obtain counter information from the registers, they are configured to monitor specific hardware events of interest. HPCs can perform thread-level profiling by saving the register values during context switches and, therefore, try to avoid contamination due to events from other processes \cite{sok_hpc_malware_detect}. With the adoption of ML techniques in security, HPCs have recently been repurposed as low-level telemetry for identifying malicious workloads in HMDs \cite{kuruvila_efficiency_hmd,sanyal_hmd_reservoir,hmd_fine_grained,demme_hpc,analyze_hmd,quant_hmd,hpc_myth_or_fact,low_barrier_hmd,explainable_hmd,ensemblehmd, hpc_ransomware, 9034972}.  

 Two main limitations of HPC-based HMDs frequently emerge in the literature: non-determinism and the limited number of monitorable HPC events. Non-determinism results from measurement errors in the HPCs \cite{sok_hpc_malware_detect}. However, the robustness of ML-agents, which can learn and generalize from broader patterns and trends in the data, potentially mitigates this. The second limitation pertains to the finite number of HPC events that can be simultaneously monitored due to limited physical registers (for instance, the Snapdragon Chipset in our study only accommodates four). This restriction limits the behavioral information available, narrowing the scope of the attack space that an HPC-based defense can cover. This constraint implies that HPC-based defenses are best suited to handle malware with prominent architectural side-channels, as recent works suggest \cite{hpc_malware_theory}. Moreover, taken together, these two limitations suggest that HPC-based HMDs which focus on a single mode of behavioral information (CPU-telemetry) coupled with non-determinism makes it vulnerable to attacks designed to skew the CPU telemetry measurements and induce mis-classifications \cite{hpc_myth_or_fact,sok_hpc_malware_detect}.

\textbf{System-wide Profiling: Dynamic Voltage and Frequency Scaling (DVFS). } DVFS is an integral part of all power management systems. It reduces the power consumption of an SoC by scaling down the voltage and frequency states of the different sub-systems (e.g., CPU, GPU, buses, caches, and memory) based on the targetted performance requirements of the software workloads. As a result, the DVFS states of a sub-system capture its activity level, providing insight into the impact of running workload on that sub-system. Security implications of the DVFS framework have been studied both from an offensive perspective \cite{clkscrew, plundervolt,biasp} and to create defenses \cite{chawla_application_inference, chawla_iotj}. Since Android OS (considered in this paper) is based on the Linux kernel, the DVFS states of the CPU and non-CPU devices are accessible through the cpufreq, and devfreq framework \cite{cpufreq,devfreq}. 

It should be noted that the DVFS channels (and other SYSFS nodes used in this work) capture the global state of the device as compared to HPCs that are used for monitoring the specific threads/processes. In mobile devices, numerous system-level threads and user applications are simultaneously contending for hardware resources. This makes the DVFS channels susceptible to noise arising from such background processes. However, in the case of mobile devices, such as the one considered in our work, Chawla et al. has empirically shown that the foreground applications predominantly influence the DVFS states of the device \cite{chawla_iotj}. The DVFS channels provide better visibility into the impact of running a  workload on the entire SoC at the cost of additional non-determinism compared to HPCs that capture a workload’s impact on the CPU core of the SoC.

\subsection{Threat Model} XMD is designed for multi-core mobile devices where a user interacts with a limited set of foreground applications (1-2) at a given time. In this model, the potential attacker can deploy malware with capabilities to manipulate application-level activities, but they do not have root access to the system. The XMD framework operates in kernel-land, as accessing the values of HPCs and system-wide telemetry channels (DVFS, SYSFS) requires root privileges. Therefore, the framework assumes the OS kernel to be secure and not compromised, an assumption shared with commercially deployed HMDs \cite{intel_tdt}. A compromised kernel, in fact, would undermine the process tracking capabilities of the Endpoint Detection and Response (EDR) system and thwart the trustworthiness of the telemetry on which the EDR relies \cite{Karantzas_2021}. Therefore, ensuring kernel-integrity is out of XMD's scope.

XMD is susceptible to collusion-based attacks where a seemingly benign, attacker-controlled thread may tamper with system-wide telemetry channels. Much like HPC-based HMDs, the non-determinism of the hardware telemetry could allow a motivated attacker to mask the activities of the malicious process amidst the noise \cite{sok_hpc_malware_detect}. However, XMD's utilization of telemetry from various non-core devices potentially provides an additional layer of defense against attackers capable of skewing measurements from a single SoC module (such as CPU telemetry). Additional limitations and potential solutions are further discussed in Section \ref{sec:discussion}.

\section{Related Works and Motivation}
Detection of malicious workloads using hardware telemetry has been extensively studied in the literature \cite{kuruvila_efficiency_hmd,sanyal_hmd_reservoir,hmd_fine_grained,demme_hpc,analyze_hmd,quant_hmd,hpc_myth_or_fact,chawla_iotj,low_barrier_hmd,explainable_hmd,ensemblehmd, hpc_ransomware, 9034972}. Due to the variabilities in the data collection methodology, we do not empirically compare against prior works but offer a qualitative comparison against the surveyed works. We categorize the drawbacks of these works as follows:

\textbf{Restricted scope of hardware telemetry collection.}
Prior works on HMD primarily focus on a single modality of data extracted from the CPU of the SoC \cite{kuruvila_efficiency_hmd,sanyal_hmd_reservoir,hmd_fine_grained,demme_hpc,analyze_hmd,quant_hmd,hpc_myth_or_fact,chawla_iotj,low_barrier_hmd,explainable_hmd,ensemblehmd, hpc_ransomware, 9034972}. These low-level signatures either contain thread-level behavior (e.g., HPC) or global behavior (e.g., CPU-DVFS). While these data-driven approaches have shown decent test accuracy on their dataset, they do not use the telemetry sources available from the different sub-systems of the SoC. They, therefore, do not realize the full potential of the HMD.

\textbf{Benchmarks used as benign workloads.} Prior works on HMD use benchmark applications for benign workloads \cite{kuruvila_efficiency_hmd,sanyal_hmd_reservoir,hmd_fine_grained,demme_hpc,analyze_hmd,chawla_iotj,explainable_hmd, hpc_ransomware}. Few works use regular benign applications (e.g., from Play Store); however, they mix these applications with benchmark applications \cite{hpc_myth_or_fact,ensemblehmd}. Compared to regular benign applications, which require interaction with the device to explore the different threads of operation, running the benchmark application is straightforward, easing the large-scale data collection process. These benchmark applications are synthetic workloads designed to test a specific functionality of the SoC and are not representative of real-world benign applications, introducing a bias in the dataset.

\textbf{Comparison against Software-based AVs.}
Prior works present a qualitative comparison of how the behavior-based detection approach of HMDs can outperform the static analysis-based detection techniques of production AV software \cite{kuruvila_efficiency_hmd,sanyal_hmd_reservoir,hmd_fine_grained,demme_hpc,analyze_hmd,quant_hmd,hpc_myth_or_fact,chawla_iotj,low_barrier_hmd,explainable_hmd,ensemblehmd, hpc_ransomware}. However, they do not present a quantitative comparison of how the performance of their proposed HMDs compares against the currently deployed production AV software.

\section{Theory of XMD}
\label{sec:theory}
We describe the hypothesis behind the design of XMD, followed by the theorem that supports the hypothesis. 
\subsection{Intuition behind XMD}
\label{ssec:hypotheses}

Zhou et al. questioned how low-level HPC telemetry, derived exclusively from a SoC's CPU, could effectively distinguish high-level behavior between benign and malicious applications \cite{hpc_myth_or_fact}. Botacin \& Gr'egio further stipulated that HPC telemetry might only encompass an attack space where malware exhibits noticeable architectural side-channel effects \cite{hpc_malware_theory}. We concur with this view, noting that an over-dependence on CPU hardware telemetry might restrict the HMD's coverage to malware with pronounced architectural side effects. We propose that accessing telemetry channels from different SoC sub-devices could widen the attack space addressed by the HMD. Providing ML agents with additional telemetry channels (e.g., low-level network telemetry) might enhance the HMD's detection performance, considering that malware often communicates with its command-and-control server.

It is important to acknowledge two fundamental assumptions implicit in our work, as well as all dynamic analysis-based approaches that rely on behavioral signatures. \textit{Assumption 1:} Different software executions affect telemetry differently. This belief sets the groundwork for our methodology, as we anticipate that benign and malicious applications will exhibit distinct behaviors in telemetry data, allowing ML models to detect these variances. \textit{Assumption 2:} Malware and benign applications are inherently different in their behavioral patterns and hence produce different signatures in telemetry data. This leads us to our central hypothesis: \textbf{An ML classifier could perform better when utilizing a more diverse set of telemetry channels from various SoC subsystems.} This hypothesis guides the design and evaluation of XMD. In the following section, we present the necessary background and definitions to construct the theorem supporting our hypothesis.

\subsection{Background and Definitions}
\begin{figure}[!hbt]
% \vspace{-5mm}
% \captionsetup{justification=centering}
    \centering
    \includegraphics[scale = 1]{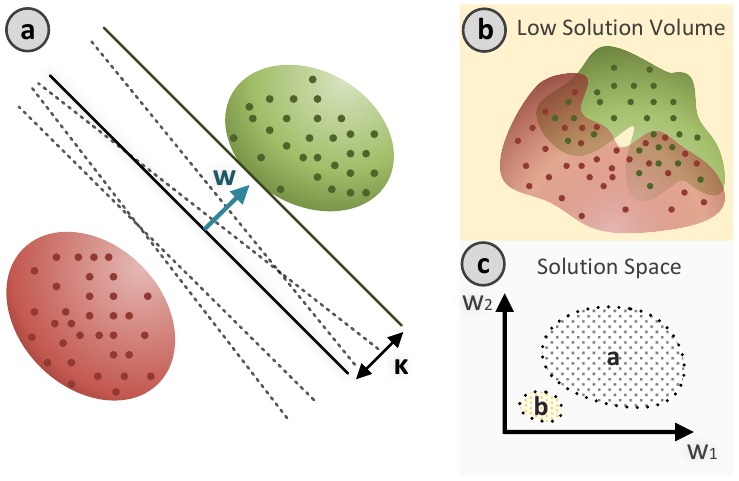}
    \caption{2-D visualization of Benign [green] vs. Malware [red] classification using manifolds: (a) candidate hyperplanes that can separate the benign and malware manifolds, (b) classification task with overlapping manifolds, (c) solution space of the hyperplanes for case-a (easier classification problem) and for case-b (difficult classification problem).}
    % \vspace{-6mm}
    \label{fig:manifold}
\end{figure}

\textbf{Manifold.} Intuitively, a manifold is a topological space that is locally Euclidean. A topological space is a set of points, with each point having its own set of neighborhoods \cite{ma2012manifold}. We have a set of features for each of the GLOBL channels and the HPC groups. Using these feature sets, we can visualize an APK sample as a representative point in a higher dimensional vector space. We construct a manifold composed of a set of these representative points for each of the benign and malware classes. Finally, we end up with a set of manifolds for each of the benign and malware classes, for all the GLOBL channels and the HPC groups. The representative points move around in their respective manifolds introducing intra-class variabilities (e.g., stochasticity). We use the notation $F^{\mu}$ with $\mu= B$ or $M$ for the benign and malware manifolds, respectively, with each point on the manifold represented by $\mathbf{x}^{\mu} \in F^{\mu}$.

A point on the manifold consists of the input space $\mathbf{x}^{\mu} \in F^{\mu}$ given as    $\mathbf{x}^{\mu}(\vec{S})=\sum_{i=1}^{D+1} S_{i} \mathbf{u}_{i}^{\mu}$ where $\mathbf{u}_{i}^{\mu}$ are a set of orthonormal bases of the $(D+1)$ dimensional linear subspace containing $F^{\mu}$, the $D+1$ components $S_{i}$ represent the coordinates of the manifold point within this subspace and are constrained to be in the set $\vec{S} \in \mathcal{S}$. $\mathcal{S}$ denotes the shape of the manifolds and encapsulates the affine constraint. 

\textbf{Separability and Hyperplanes. } The goal of the classification task is to learn the position of the decision hyperplane between the two object manifolds. The ability of a classifier to discriminate between two class manifolds can be mapped to the separability of class manifolds by a linear hyperplane \cite{dicarlo2007untangling, chung2018classification}. We study the separability of the benign and malicious manifolds into separate classes, denoted by binary labels $y^{\mu}=\pm 1$, by a linear hyperplane. As shown in Figure \ref{fig:manifold}.(a), a hyperplane is described by a weight vector $\mathbf{w} \in \mathbb{R}^{N}$, and separates the manifolds with margin $\kappa$ such that
    $y^{\mu} \mathbf{w} \cdot \mathbf{x}^{\mu} \geq \kappa$
for all $\mu$ and $\mathbf{x}^{\mu} \in F^{\mu}$. 

\textbf{Solution Volume. } We use solution volume $\mathcal{V}$ as a metric to characterize the separability of two manifolds \cite{gardner1988space}. A higher solution volume results in better and more generalizable classification. An intuitive representation of the solution volume is shown in Figure \ref{fig:manifold}.(c). We can observe a higher solution volume for case-a, where the benign and malware manifolds have higher separability, than case-b, where the manifolds overlap, resulting in lower separability and, therefore, a lower solution volume. Following Gardner's replica framework \cite{gardner1988space}, the volume $\mathcal{V}$ of the solution space is defined as
\begin{equation}
    \mathcal{V}=\int d^{N} \mathbf{w} \delta\left(\|\mathbf{w}\|^{2}-N\right) \prod\limits_{\mu, \mathbf{x}^{\mu} \in F^{\mu}} \Theta\left(y^{\mu} \mathbf{w} \cdot \mathbf{x}^{\mu}-\kappa\right)
    \label{eq:vol}
\end{equation}

where $\Theta(\cdot)$ is the Heaviside function to enforce the margin constraints in the linear separation constraint $y^{\mu} \mathbf{w} \cdot \mathbf{x}^{\mu} \ge \kappa$, along with the delta function to ensure $\|\mathbf{w}\|^{2}=N$.

\textbf{Convex Hull and Polytopes:} The convex hull of a set of points $\mathcal{S}$ is the intersection of all half-spaces that contain $\mathcal{S}$. A half-space is either of the two parts into which a hyperplane divides an affine space. For example, in a two-dimensional Euclidean space, a half-space is either of the two parts into which the space is divided by a line. A convex polytope is an intersection of a finite number of half-spaces. 

Mathematically, the convex hull is given as: $\mathcal{CH}\left(F^{\mu}\right)=\left\{\mathbf{x}^{\mu}(\vec{S}) \mid \vec{S} \in \mathcal{CH}(\mathcal{S})\right\}$, where
\begin{equation}
    \mathcal{CH}(\mathcal{S})=\left\{\sum_{i=1}^{D+1} \alpha_{i} \vec{S}_{i} \mid \vec{S}_{i} \in \mathcal{S}, \alpha_{i} \geq 0, \sum_{i=1}^{D+1} \alpha_{i}=1\right\}
\end{equation}

\subsection{Theorem}
\label{ssec:theorems}
Using the formal theory of linear separability of the object manifolds \cite{chung2018classification}, we perform an analytical study that supports our hypothesis presented in Section \ref{ssec:hypotheses}. The theorem establishes that XMD's superior performance stems from a higher solution volume which arises from the fusion of information of telemetry channels from different devices of an SoC, e.g., CPU, GPU, Network, Memory, and Cache.

\noindent \textbf{Theorem 1: }
\textit{Let $\mathcal{V}_i$ be the solution volume corresponding to the classification task of the benign and malware applications using the $i$-th telemetry channel in the $N$-dimensional vector space, where the $i$-th basis corresponds to the $i$-th telemetry channel $\forall i \in [1, N]$, and $N$ is the total number of telemetry channels. We show that the solution volume arising from the union of different $\mathcal{V}_i$s, \textit{i.e.} $\left(\bigcup_{i} \mathcal{V}_i\right)$,  is greater than the individual $\mathcal{V}_i$s considered independently.}

To prove this, we assume an N-dimensional vector space, with one basis for each of the N telemetry channels, and each $\mathcal{V}_i$ is an orthogonal projection of the union of solution volumes $\left(\bigcup_{i} \mathcal{V}_i\right)$ on the i-th basis. Next, we show that the solution volume arising from $\bigcup_{i} \mathcal{V}_i$ is lower bounded by the maximum solution volume $\mathcal{V}_{\max}$, where $\mathcal{V}_{\max}$ is $\max\{\mathcal{V}_i\ \ \forall i\in[1,N]\}$ . A higher solution volume results in better classification performance, hence, supports our hypothesis.
 
\subsection{Mathematical proof}
\label{appendix:math_proof}
\par \textit{\textbf{Lemma 1:}} \begin{equation}
\begin{aligned}
    \mathcal{V}[\mathcal{CH}\left(\bigcup_{i} \mathcal{V}_i\right)]
    \ge \max\{\mathcal{V}_i\}
\end{aligned}
\end{equation}

\textbf{\textit{Proof: }} 
Let $\mathcal{V}_i$ be nonempty, convex sets. We show that $x \in \mathcal{CH}(\bigcup_{i} \mathcal{V}_{i})$ if and only if there exist elements $v_{i} \in \mathcal{V}_{i}$ and $\lambda_{i} \geq 0$  with $\sum_{i} \lambda_{i}=1$ such that $x=\sum_{i} \lambda_{i} v_{i}.$ This can be represented as
$$
\mathcal{CH}(\bigcup_{i} \mathcal{V}_{i})=\left\{\sum_{i} \lambda_{i} v_{i} \mid \sum_{i} \lambda_{i}=1, \lambda_{i} \geq 0, v_{i} \in \mathcal{V}_{i} \right\}
$$
Now, let us consider the solution volume $\mathcal{V}$ as a measure defined on a vector space $\Omega$. Then we show that $\mathcal{V}(A) \leq \mathcal{V}(B)$ for all $A \subset B \in \Omega$. Let $A \subset B$, let $C=A^{c} \cap B$. Then $A \cap C=\emptyset$ and $A \cup C=B$. Thus, $\mathcal{V}(A \cup C)=\mathcal{V}(A)+ \mathcal{V}(C)=\mathcal{V}(B)$. Therefore $\mathcal{V}(A) \leq \mathcal{V}(B)$ for $\mathcal{V}(C) \geq 0$ by non-negativity.
Hence, the solution volume of a convex manifold $\mathbf{X}$, given as $\mathcal{V}\{\mathbf{X}\}$, is a monotonic increasing measure. Therefore,
\begin{align} 
[\mathcal{V}\{\mathcal{CH}\left(\bigcup_{i} \mathcal{V}_i\right)\}]  &=  [\mathcal{V}\{\mathcal{CH}\left(\mathcal{V}_{\max} \cup \mathcal{V}_i\backslash\mathcal{V}_{\max}\right)\}] \nonumber \\
&\ge [\mathcal{V}_{\max} \cup \mathcal{V}_i\backslash\mathcal{V}_{\max}]  \nonumber \\
 &\ge  [\mathcal{V}_{\max}]  =  \max\{[\mathcal{V}_i ] \quad \forall i\} \nonumber \\
\Rightarrow \mathcal{V}[\mathcal{CH}\left(\bigcup_{i} \mathcal{V}_i\right) ] 
    &\ge \max\{(\mathcal{V}_i)\} \end{align}$\blacksquare$

\par \textbf{Theorem 1: } Let $\mathcal{V}_i$ be the solution volume corresponding to the telemetry channel-$i$ in the $N$-dimensional vector space where the $i$-th basis corresponds to the $i$-th telemetry channel $\forall i \in [1, N]$, and $N$ is the total number of telemetry channels. Then, $\mathcal{V}[\mathcal{CH}\left(\bigcup_{i} \mathcal{V}_i\right)] \ge \max\{\mathcal{V}_i\} \quad \forall i  \in [1, N]$ .
 
\textbf{Short Proof:} 
 We consider a convex polytope for each of the solution volume $\mathcal{V}_i$. Without loss of generality, we assume that each $\mathcal{V}_i$ is an orthogonal projection of the union of the solution volumes ($\bigcup_{i} \mathcal{V}_i$), which we refer to as the universal convex polytope. The universal convex polytope is constructed by taking a convex hull over the union of its $N$ orthogonal components $\mathcal{V}_i$. From Lemma 1, we get
    $\mathcal{V}[\mathcal{CH}\left(\bigcup_{i} \mathcal{V}_i\right)] \ge \max\{\mathcal{V}_i\}$.
$\blacksquare$

 \begin{figure*}[!ht]
% \captionsetup{justification=justifying}
% \vspace{-5mm}
    \centering
    \includegraphics[scale = 0.85]{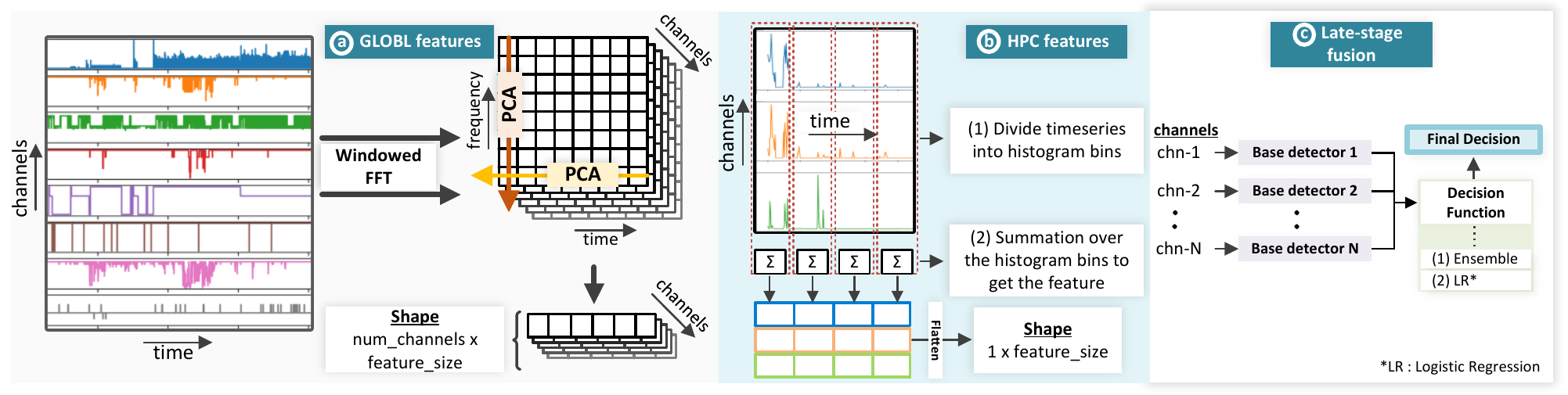}
    \caption{Creating the classifiers: (a) Feature engineering steps for GLOBL channels, (b) feature engineering steps for the HPC groups, (c) Late stage fusion for merging the decisions of the base detectors}
    % \vspace{-6mm}
    \label{fig:feature_engineering}
\end{figure*}

\begin{table*}[!htbp]
\centering
\caption{Malware characterization using AVClass \cite{avclass2}}
\label{tab:malware_char}
\resizebox{0.7\textwidth}{!}{%
\begin{tabular}{|cc|cc|cc|cc|}
\hline
\textbf{Class} & \textbf{\#apks} & \textbf{Family} & \textbf{\#apks} & \textbf{Family} & \textbf{\#apks} & \textbf{Family} & \textbf{\#apks} \\ \hline
grayware   & 457 & hiddad     & 47 & piom            & 5 & autoins           & 2 \\
adware     & 159 & dnotua     & 33 & blacklister     & 4 & iconhider         & 2 \\
downloader & 40  & appsgeyser & 24 & marsdaemon      & 4 & gamclk            & 2 \\
virus      & 23  & joker      & 21 & datacollector   & 3 & apptrack          & 1 \\
tool       & 19  & fakeapp    & 15 & silentinstaller & 3 & wapron            & 1 \\
clicker    & 6   & smsreg     & 8  & umpay           & 2 & baiduprotect      & 1 \\
rogueware  & 1   & smspay     & 6  & gapussin        & 2 & contactscollector & 1 \\ \hline
\end{tabular}%
}
\end{table*}

\subsection{Approach for Experimental Validation}
To validate the hypothesis and theorem presented in Section \ref{ssec:hypotheses} and \ref{ssec:theorems} respectively, we design a robust data collection framework that collects the two classes of telemetry (HPC and GLOBL) simultaneously and collectively captures a workload's impact on different sub-systems of the SoC like CPU, GPU, memory, buses, and network (Section \ref{sec:data_collection_framework}).  The framework incorporates measures to reduce the datasets' bias and prevent over-optimistic results. For example, we devise a Logcat-based activation checker that filters out the runs in which the application does not have sufficient runtime. For both the HPC data logs and the GLOBL data logs, the data collection process produces multi-variate time series data. The feature engineering choices are the same as the ones used in prior works for DVFS \cite{chawla_iotj, chawla_application_inference}  and HPC \cite{demme_hpc, hpc_myth_or_fact}, and are summarized in Figure \ref{fig:feature_engineering}.(a) and \ref{fig:feature_engineering}.(b). Using the fusion-based approach summarized in Figure \ref{fig:feature_engineering}.(c), we demonstrate that incorporating the expansive telemetry improves the classification performance of the fusion-based model used in XMD, validating Theorem 1.

\section{Dataset Collection Framework}
\label{sec:data_collection_framework}

\subsection{Malware and Benign programs used}

\textbf{STD-DATASET}: In this paper, we use a broad definition of malware, i.e., any application that has been flagged as malicious by at least one AV on VirusTotal \cite{calvallaro_piggybacking}. The dataset consists of real-world Android benign and malware applications. In particular, we acquired an initial dataset of 1033 samples of malicious and 723 samples of benign Android applications. The samples were downloaded from AndroZoo and were collected in the period from Dec 2019 to June 2021 \cite{androzoo}. There are 54 malware families in the malware dataset, as reported by the ESET-NOD32 AV engine from VirusTotal. The number of applications and the total number of iterations is summarized in Table \ref{tab:dataset_taxanomy}.

To gain a deeper understanding of the malware in our dataset, we employed the AVClass tool \cite{avclass2} to analyze the Virustotal report generated for our malware samples. The results of this analysis are presented in Table \ref{tab:malware_char}. This table classifies the malware samples into different categories, based on class and family, and lists the number of samples that correspond to each specific class or family. Notably, these tags are not exclusive, meaning a single sample might be classified as both `grayware' and `adware'. Majority of the samples received the `grayware' tag. Previous research by Chen et al. on characterization of Android malware indicated that grayware could embody an array of potentially harmful functionalities \cite{grayware}. In our dataset, we have observed such functionalities, as seen in categories like Redirected Promotion Apps (hiddad), Data Collection Apps (dnotua, joker), Fake apps (fakeapp), and Dialing/SMS-Managing Apps (smsreg, smspay) \cite{grayware}. \textit{Many of these applications may not cause detectable architectural side-channel effects, suggesting that they could evade detection by HPC-telemetry based HMDs.}  Thus, our dataset could pose a robust test for the expanded attack space covered by our proposed method, underlining our argument that the inclusion of a wider array of telemetry channels could improve malware detection capabilities.

\subsection{Bare-metal analysis environment}
\label{ssec:environment_description}
We built a host-client-based bare-metal sandbox environment to perform a large-scale automated collection of hardware-level telemetry. The Client is a Google Pixel 3 mobile device (running Android OS) on which the benign and malicious samples are executed, and the host is a Linux OS-based PC that orchestrates the data collection. During application execution, we simulate human interaction using the Monkey tool \cite{monkey} and Broadcast Events \cite{dissect_android_malware} to trigger the malware payloads. We leveraged Android Debug Bridge (\verb|adb|) \cite{adb} for transferring data and sending commands between the host and the Client. The Client is connected to the Internet, which is crucial for the malware to perform essential functionalities like communicating with its command-and-control (C2) server. Since we were using a bare-metal analysis environment, we developed a custom checkpointing scheme to restore the Client's OS to a clean state between the application runs.

\textbf{Environment assumptions.} During data collection of the foreground application, multiple system workloads (e.g., \textit{system\_server}, \textit{surfaceflinger}, \textit{servicemanager}, \textit{mediaserver}, etc.) were executing in the background. Some of these workloads (e.g., \textit{system\_server}) were sending and receiving data packets on the network. Since the experimental Android device has a multi-core CPU, the processes belonging to the foreground application were context switching in and out of the cores and migrating to different cores throughout data collection. As indicated by prior research, this introduces measurement errors and non-determinism \cite{sok_hpc_malware_detect}. While previous works have attempted to pin the process to a core during data collection to reduce noise from measurement errors \cite{hpc_myth_or_fact, demme_hpc}, we contend that this approach is not realistic. HMD is intended to function as a real-time detector that continuously runs in the background. Post-deployment, HMDs will profile processes subject to context switching, resulting in inherently erroneous measurements. Therefore, during data collection, we do not pin the target process to any particular core. We acknowledge that this makes the problem more difficult (but realistic) and necessitates the design of effective Machine Learning (ML) agents that can generalize beyond the measurement errors introduced during the data-collection process. Overall, we ensure that the data collection environment is similar to the environments observed in a scenario where a Android user is interacting with one foreground application with multiple system workloads running in the background.

\subsection{Selection of Hardware Signals}
\begin{table}[!htbp]
\centering
    \captionof{table}{List of collected HPC events}\label{tab:hpc-events} 
    \resizebox{0.47\textwidth}{!}{%
    \begin{tabular}{|c|c|c|}
    \hline
    \textbf{Group}                    & \textbf{HPC event}  & \textbf{Reference} \\ \hline
    \multirow{3}{*}{\textit{group-1}} & cpu-cycles          & \cite{explainable_hmd}                  \\ \cline{2-3} 
                                      & instructions        & \cite{kuruvila_efficiency_hmd, hmd_fine_grained, analyze_hmd, quant_hmd, low_barrier_hmd, hpc_ransomware}                  \\ \cline{2-3} 
                                      & raw-bus-accesses    & \cite{demme_hpc, quant_hmd}                 \\ \hline
    \multirow{3}{*}{\textit{group-2}} & branch-instructions & \cite{kuruvila_efficiency_hmd, hmd_fine_grained, demme_hpc, analyze_hmd, low_barrier_hmd, explainable_hmd, hpc_ransomware}                  \\ \cline{2-3} 
                                      & branch-misses       & \cite{explainable_hmd, sanyal_hmd_reservoir, hmd_fine_grained, low_barrier_hmd, explainable_hmd, hpc_ransomware}                  \\ \cline{2-3} 
                                      & raw-mem-access      & \cite{ensemblehmd, demme_hpc, quant_hmd}                  \\ \hline
    \multirow{3}{*}{\textit{group-3}} & cache-references    & \cite{low_barrier_hmd, hpc_myth_or_fact, hpc_ransomware}                  \\ \cline{2-3} 
                                      & cache-misses        & \cite{low_barrier_hmd,kuruvila_efficiency_hmd, sanyal_hmd_reservoir, hmd_fine_grained, hpc_ransomware}                  \\ \cline{2-3} 
                                      & raw-crypto-spec     & -                  \\ \hline
    \multirow{3}{*}{\textit{group-4}} & bus-cycles          & \cite{sanyal_hmd_reservoir, explainable_hmd}                  \\ \cline{2-3} 
                                      & raw-mem-access-rd   & \cite{demme_hpc, quant_hmd}                  \\ \cline{2-3} 
                                      & raw-mem-access-wr   & \cite{demme_hpc, quant_hmd}                  \\ \hline
    \end{tabular}%
    }

\end{table}

\begin{table*}[!htbp]
\centering
\caption{List of collected GLOBL (DVFS \& SYSFS) channels.}
\label{tab:dvfs_sysfs_channels}
\resizebox{0.8\textwidth}{!}{%
\begin{tabular}{c|c|l|l|}
\cline{2-4}
\multicolumn{1}{l|}{} & \textbf{Channel Number} & \multicolumn{1}{c|}{\textbf{Location}} & \multicolumn{1}{c|}{\textbf{Description}} \\ \hline
\multicolumn{1}{|c|}{\multirow{11}{*}{\textbf{\begin{tabular}[c]{@{}c@{}}D\\ V\\ F\\ S\end{tabular}}}} & 1 & /sys/class/devfreq/5000000.qcom,kgsl-3d0/gpu\_load & GPU controller \\ \cline{2-4} 
\multicolumn{1}{|c|}{} & 2 & /sys/devices/system/cpu/cpu0/cpufreq/scaling\_cur\_freq & CPU controller : lower cluster \\ \cline{2-4} 
\multicolumn{1}{|c|}{} & 3 & /sys/devices/system/cpu/cpu7/cpufreq/scaling\_cur\_freq & CPU controller : higher cluster \\ \cline{2-4} 
\multicolumn{1}{|c|}{} & 4 & /sys/class/devfreq/soc:qcom,cpubw/cur\_freq & CPU bus bandwidth controller \\ \cline{2-4} 
\multicolumn{1}{|c|}{} & 5 & /sys/class/devfreq/soc:qcom,gpubw/cur\_freq & GPU bus bandwidth controller \\ \cline{2-4} 
\multicolumn{1}{|c|}{} & 6 & /sys/class/devfreq/soc:qcom,kgsl-busmon/cur-freq & GPU bus bandwidth controller \\ \cline{2-4} 
\multicolumn{1}{|c|}{} & 7 & /sys/class/devfreq/soc:qcom,l3-cpu0/cur-freq & Latency controller L3 cache : lower cluster \\ \cline{2-4} 
\multicolumn{1}{|c|}{} & 8 & /sys/class/devfreq/soc:qcom,l3-cpu4/cur-freq & Latency controller L3 cache : higher cluster \\ \cline{2-4} 
\multicolumn{1}{|c|}{} & 9 & /sys/class/devfreq/soc:qcom,llccbw/cur-freq & Last level cache controller \\ \cline{2-4} 
\multicolumn{1}{|c|}{} & 10 & /sys/class/devfreq/soc:qcom,memlat-cpu0/cur-freq & Memory latency controller : lower cluster \\ \cline{2-4} 
\multicolumn{1}{|c|}{} & 11 & /sys/class/devfreq/soc:qcom,memlat-cpu4/cur-freq & Memory latency controller : higher cluster \\ \hline
\multicolumn{1}{|c|}{\multirow{4}{*}{\textbf{\begin{tabular}[c]{@{}c@{}}SYS\\ FS\end{tabular}}}} & 12 & /sys/class/net/tun0/statistics/rx\_bytes & Network : received bytes \\ \cline{2-4} 
\multicolumn{1}{|c|}{} & 13 & /sys/class/net/tun0/statistics/tx\_bytes & Network : transmitted bytes \\ \cline{2-4} 
\multicolumn{1}{|c|}{} & 14 & /sys/class/power\_supply/battery/current\_now & Device current \\ \cline{2-4} 
\multicolumn{1}{|c|}{} & 15 & /sys/class/power\_supply/battery/voltage\_now & Device voltage \\ \hline
\end{tabular}%
}
\end{table*}

\textbf{Choice of HPC channels. }
The HPC events are stated in Table \ref{tab:hpc-events}. We performed a comprehensive literature survey to identify the performance counter events used in prior work that resulted in the best detection performance. We consider an additional HPC event, not used in prior work, called \verb |raw-crypto-spec|, which can potentially capture essential malware functionalities (SSL or TLS handshake). The number of HPC available on the device limits the collection of HPC events. On the Snapdragon chipset, there are four available HPCs. However, one HPC is repurposed for monitoring memory latency, so we can collect at most three events simultaneously. Therefore, all the HPC events are divided into groups 1-4. Each HPC group has three events that are collected simultaneously in a single iteration.

\textbf{Choice of DVFS channels.}
Prior works have primarily focused on the impact of running a workload on the CPU by monitoring the DVFS states of the CPU controller \cite{chawla_application_inference,chawla_iotj}; we expand on this notion by considering an expansive set of DVFS channels that cover both the CPU and the non-CPU devices like GPU, memory, buses, and caches. Channel-1 to Channel-11 in Table \ref{tab:dvfs_sysfs_channels} elaborates on the selected DVFS channels, their corresponding locations in the Linux device tree, and the device's signature they capture.

\textbf{Choice of SYSFS channels. }
To capture the low-level impact of benign or malicious workload on Network devices, we recorded the number of bytes transmitted and received by the device. These channels are Channel-12 and Channel-13 in Table \ref{tab:dvfs_sysfs_channels}. Prior works have demonstrated the efficacy of power side channels to detect malicious workloads on x86, IoT, and embedded systems \cite{power_malware1, power_malware2, power_malware_embedded,deeppower}.
We collected telemetry from the \verb |voltage_now| (Channel-14) and \verb|current_now| (Channel-15) sysfs nodes. While the channels and events described earlier capture the impact of workload on a specific subsystem of the SoC, the power channels present a global telemetry channel capturing the impact on the entire SoC.

Overall, the combined information from the HPC, DVFS, and SYSFS channels present a comprehensive low-level behavior over all the sub-modules of an SoC, like the CPU, caches, GPU, memory, buses, and network.

\begin{figure}[!htb]
    % \captionsetup{justification=justifying}
    \centering
    \includegraphics[scale = 0.5]{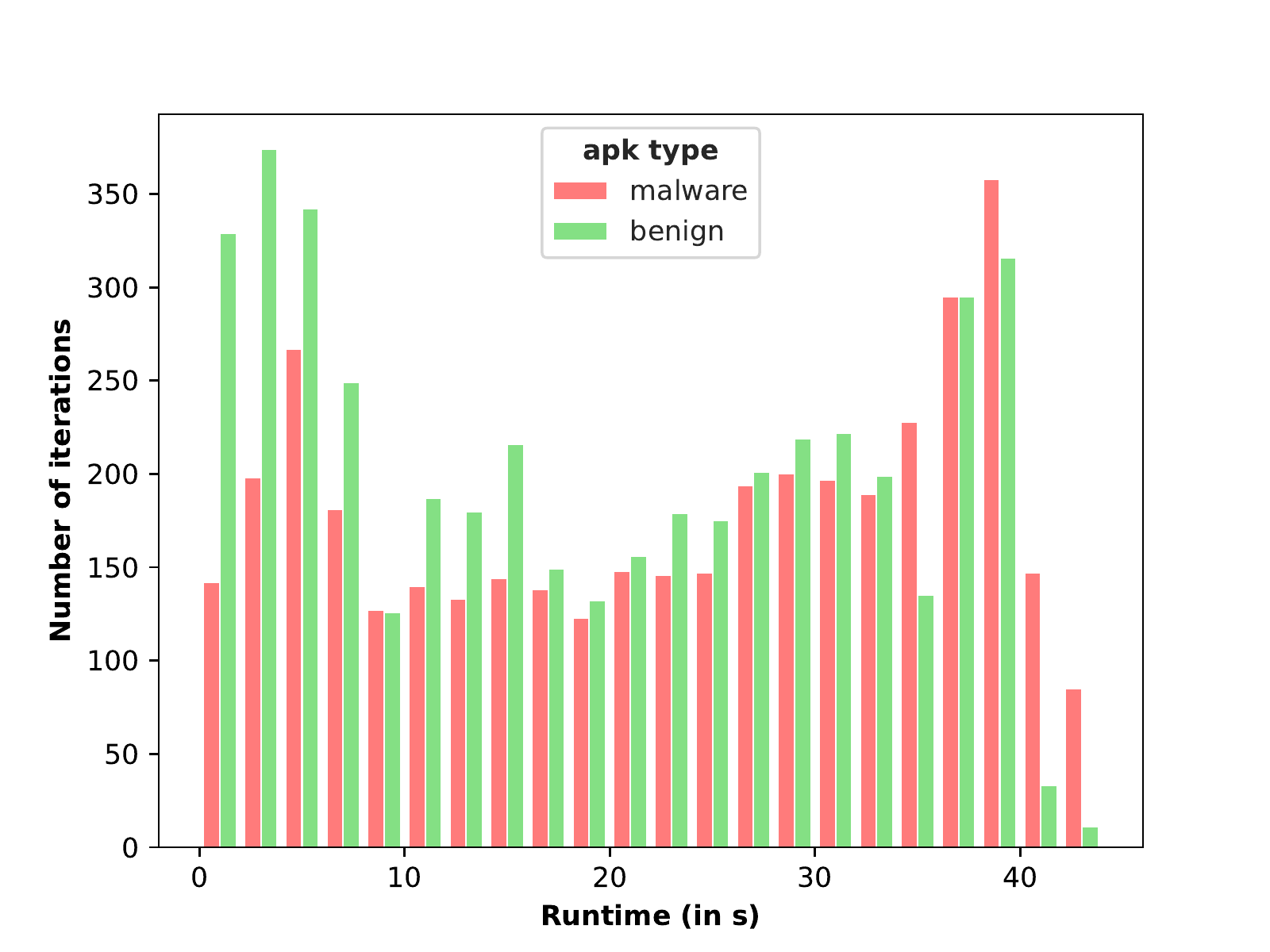}
    \caption{Runtime for different iterations of malware and benign applications (using Logcat)}
    \label{fig:runtime_benign_malware}
\end{figure}

\begin{table*}[!htb]
\centering
\caption{Details of the dataset}
\label{tab:dataset_taxanomy}
\begin{tabular}{|c|c|c|c|ccccc|}
\hline
\multirow{3}{*}{\textbf{Application}} &
  \multirow{3}{*}{\textbf{\# apk}} &
  \multirow{3}{*}{\textbf{\begin{tabular}[c]{@{}c@{}}\# apk \\ executed once\\ (\% activation)\end{tabular}}} &
  \multirow{3}{*}{\textbf{\begin{tabular}[c]{@{}c@{}}\# apk\\ post logcat filter\\ (\% activation)\end{tabular}}} &
  \multicolumn{5}{c|}{\textbf{\# Files}} \\ \cline{5-9} 
 &
   &
   &
   &
  \multicolumn{1}{c|}{\multirow{2}{*}{\textbf{GLOBL}}} &
  \multicolumn{4}{c|}{\textbf{HPC}} \\ \cline{6-9} 
 &
   &
   &
   &
  \multicolumn{1}{c|}{} &
  \multicolumn{1}{c|}{\textbf{group-1}} &
  \multicolumn{1}{c|}{\textbf{group-2}} &
  \multicolumn{1}{c|}{\textbf{group-3}} &
  \textbf{group-4} \\ \hline
Benign &
  723 &
  681 (94\%) &
  448 (62\%) &
  \multicolumn{1}{c|}{2120} &
  \multicolumn{1}{c|}{408} &
  \multicolumn{1}{c|}{602} &
  \multicolumn{1}{c|}{582} &
  582 \\ \hline
Malware &
  1033 &
  776 (75\%) &
  555 (54\%) &
  \multicolumn{1}{c|}{2143} &
  \multicolumn{1}{c|}{577} &
  \multicolumn{1}{c|}{637} &
  \multicolumn{1}{c|}{578} &
  499 \\ \hline
\end{tabular}
\end{table*}

\subsection{Methodology}
\label{ssec:data_collection_architecture}

\textbf{Method for Running Experiments.}
We perform eight independent iterations of data collection for each Android application, where a different sequence of interactions was used for each iteration. For each iteration of data collection, we collected all the GLOBL channels, and one group of HPC events since the number of HPC registers limits us. Overall, each Android application has eight iterations of data logs for the GLOBL channels and two iterations of data logs for each HPC group. We restrict data collection for each iteration to 40 seconds for each iteration, a decision informed by space and time considerations.

\textbf{Rationale behind 40-second execution time for ML model training and evaluation.} Recent research by Kuchler et al. has investigated the impact of execution time on the quality of behavioral telemetry (system calls) and subsequent classification performance of ML model \cite{sandboxExecution}. Their findings highlight that extending the data collection window from one to two minutes marginally influences the performance of the ML classifier, with ROC-AUC scores of 0.968 and 0.967, respectively. Therefore, the most valuable behavioral telemetry from malware samples is collected within the first minute; thus, extending the execution window does not necessarily yield more meaningful features for training the ML model. Given these findings, our methodology has been designed to collect data for 40 seconds per iteration. Repeated across eight independent iterations for each application, this accumulates 320 seconds of GLOBL telemetry and 80 seconds of HPC telemetry per HPC group. This aligns our method with Kuchler et al.'s insights and ensures ample informative behavioral telemetry is available for training the ML model. For evaluation, we recognize that the change in the predictive performance of the ML model due to non-core telemetry augmentation could vary across execution time frames. To examine this, we perform sensitivity analysis using time frame increments from 5 to 40 seconds, detailed in Section \ref{ssec:hypothesis_vs_exec_time}. Overall, the chosen 40-second time frame facilitates effective training and evaluation of the ML model, focusing on testing the central hypotheses of our study. However, we acknowledge that our study does not include a comprehensive long-term evaluation of 90-120s, which is a direction for future exploration (Section \ref{sec:discussion}).

\textbf{Dataset split.} We divide the dataset into three splits: train (70\%), trainSG (15\%), and test (15\%). We train the base-classifiers for GLOBL-channels and HPC-groups using the train split. We train the Stacked Generalization models for the late-stage decision-fusion from base-classifiers using the trainSG split. The final scores are reported using the test split.

\textbf{Monitoring execution using Logcat :} 
In Android OS, logs from applications are collected in a series of circular buffers, which can be filtered and viewed using Logcat \cite{logcat}. The logs contain the timestamp of the activity, the PID, and the description of the activity. We designed a Logcat-based activation checker to ensure that the malware is executing in the foreground while the hardware-telemetry logs are collected in the background. For every iteration of data collection, we collected its corresponding logcat and used it to calculate the execution time of the foreground application. 

Figure \ref{fig:runtime_benign_malware} shows the distribution of the runtimes of different iterations of the benign and malware applications, calculated using the Logcat logs. We perform a grid search to identify the logcat-based filter's threshold that maximizes the classifier's predictive performance. Therefore, for this work, the logcat-based filter rejected all the iterations in which the foreground application executed for less than 15 seconds. Table \ref{tab:dataset_taxanomy} summarizes the number of files post-filtering for the GLOBL channels and each HPC group. E.g., we observe that 75\% of malware executed in at least one of the eight iterations, and 54\% of malware cleared the logcat filter. On the other hand, benign applications have a higher activation rate, with 95\% of applications executing at least once and 62\% of applications clearing the logcat filter.

\section{Analysis}
\begin{figure}[!htb]
\captionsetup{justification=centering}
% \vspace{-5mm}
    \centering
    \includegraphics[scale = 0.33]{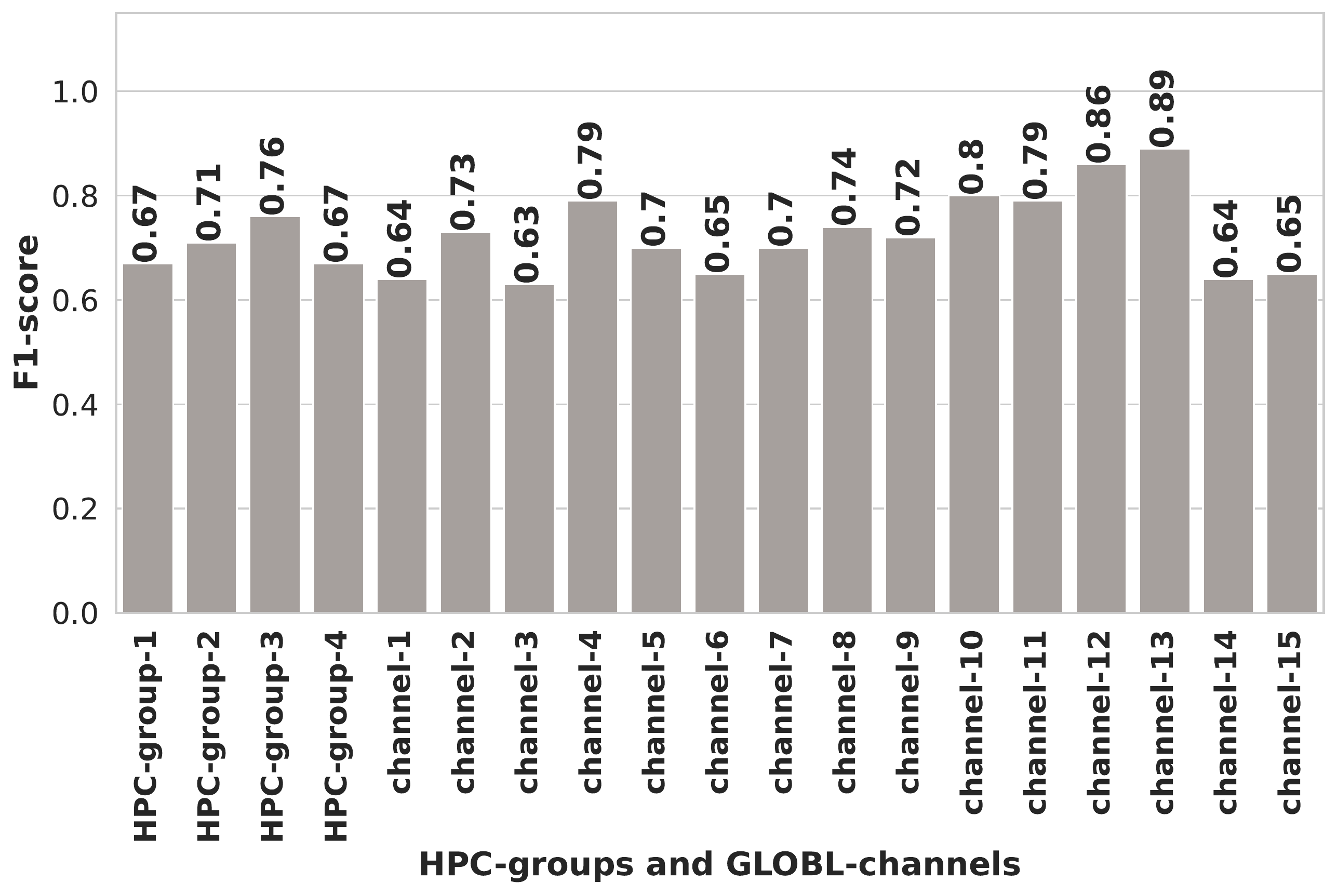}
    \caption{F1-score for HPC-groups and GLOBL channels}
    \label{fig:dvfs_benign_benchmark_malware}
\end{figure}

In this section, we perform empirical analysis aimed at validating the theorem. Using \textsc{std-dataset}, we first demonstrate that the solution volumes of the different telemetry channels are distinct. Finally, we demonstrate that fusing the solution volumes of the different channels results in increased separability of the benign and malware class and improves the predictive performance of the classifier.

\subsection{Characterizing telemetry channels.}
\label{ssec:characterize_telemetry}
 Prior works on HMD have reported their best results using Random Forest (RF) classifiers; hence we only consider the RF classifiers for creating base-classifiers for each of the GLOBL channels and the HPC groups. Figure \ref{fig:dvfs_benign_benchmark_malware} presents the classification performance of the different HPC-based classifiers and the GLOBL-based classifiers on the \textsc{std-dataset}. We provide a brief discussion highlighting the key results and takeaways.

The F1-scores for the HPC groups and CPU-telemetry channels (GLOBL channel-2 and 3) are lower than what has been reported in prior works that have used benchmark applications labeled as benign \cite{kuruvila_efficiency_hmd,sanyal_hmd_reservoir,hmd_fine_grained,demme_hpc,analyze_hmd,chawla_iotj,explainable_hmd}. Using real-world benign applications results in a more realistic scenario than using benchmark applications, making the classification task difficult. The HPC-groups do not have the performance advantage over the GLOBL channels, signifying that CPU-telemetry is insufficient to separate malicious applications from real-world benign applications, despite the accurate profiling power resulting from the thread-level profiling. 

The GLOBL channels-12 and 13 that capture the number of transmitted and received bytes offer significantly higher F1-score compared to the other channels and HPC groups, demonstrating the role of communication with the C2 server in differentiating a malicious from a benign workload. Despite the higher F1-score, relying solely on low-level telemetry from one subsystem (e.g., network) creates a single mode of failure that can be easy to bypass \cite{sok_hpc_malware_detect}.

There is a wide variation in the F1-scores (0.63-0.89) of ML models using the same learning algorithm for the different telemetry channels. This indicates that the solution volume $\mathcal{V}$ of different telemetry channels are distinct, potentially arising due to the different sources of information captured by the telemetry channels. 

\begin{mdframed}[style=mpdframe,frametitle=Takeaway-1]
The solution volume $\mathcal{V}$ of different telemetry channels are distinct.
\end{mdframed}

\subsection{Validating Theorem-1: Is fusion-based model the right approach?}
\label{sssec:t_test}

In the previous section, we observed that each telemetry class has strengths and weaknesses. HPCs can offer thread-level profiling but can only profile the CPU. GLOBL channels capture a global impact of running a workload on the SoC but cannot identify the malicious threads. This motivates us to develop a fusion-based approach called XMD that complements the thread-level profiling provided by the HPCs with the global profiling provided by the GLOBL channels.

\textit{Theorem-1} in Section \ref{sec:theory} guides the design of a fusion-based model where using multiple telemetry channels results in better solution volume than considering a single telemetry source, e.g., the CPU-telemetry-based classifiers. We study the impact of incorporating more telemetry channels on the distinguishability of the benign and malware applications using Dissimilarity Scores (DS) derived from the Welch t-test. Welch's t-test is a statistical measure to quantify the similarity between two populations (e.g., benign and malware) using their average statistics. It is a type of hypothesis testing where the null hypothesis is accepted or rejected based on the calculated t-statistic and its corresponding p-value. The null hypothesis for this study is: ``\textit{hardware telemetry signatures of malware and benign applications have similar means}." The t-statistics are calculated using Equation \ref{eqn:t_stat}.

\begin{equation}
    \label{eqn:t_stat}
    t\text{-}statistic=\frac{\mu_{1}-\mu_{2}}{\sqrt{\frac{{s_{1}}^2}{N_{1}} + \frac{{s_{2}}^2}{N_{2}}}}
    \end{equation}

\noindent where $\mu_{1}$ and $\mu_{2}$ are the mean of the sample under observation, $s_{1}$ and $s_{2}$ are the variance with $N_{1}$ and $N_{2}$ being the total number of samples. When the calculated $t-statistic > |4.5|$, we can reject the null-hypothesis with p-value of $1$x$10^{-5}$ and confidence score of 99.999\% \cite{t_test}.

We perform pairwise t-test analysis on post-processed features. We use the t-statistics to estimate a dissimilarity score (DS) given by Equation \ref{eqn:ds_eqn}. 

\begin{equation}
    \label{eqn:ds_eqn}
    DS= \frac{card({\{f_{i}\ :\ t\text{-}statistic_{f_{i}}\;>\;|4.5|\}})}{|\mathbf{F}|}
    \end{equation}

\noindent where $card(.)$ denotes the cardinality of the set, $f_{i}$ is the $i_{th}$ component of the feature vector $\mathbf{F}$, and $|.|$ is dimension of the corresponding vector space. Intuitively, DS calculates the fraction of features in the feature vector where the null hypothesis is rejected, i.e., the telemetry of the malware and benign applications are distinguishable. A higher DS implies more distinguishability between malware and benign samples.

\begin{figure}[!ht]
% \vspace{-5mm}
    \includegraphics[scale = 0.3]{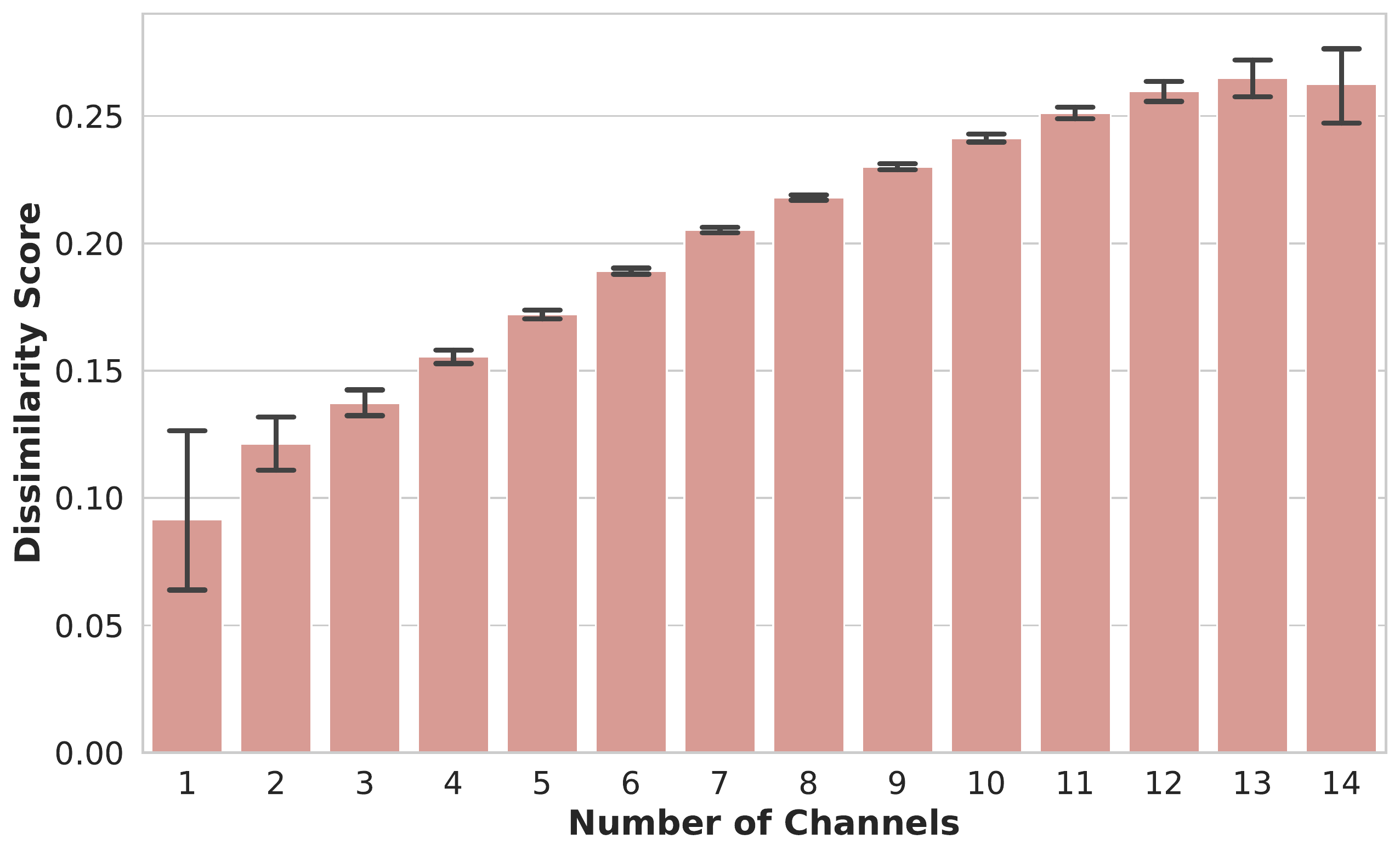}
    \caption{DS score vs. Number of channels indicating increased distinguishability between malware and benign classes as more channels are added.}
    % \vspace{-6mm}
    \label{fig:ds_score_pic}
\end{figure}

For each pair of benign and malware applications, we incorporate more telemetry channels into the feature vector. This is followed by Principal Component Analysis to reduce the augmented feature size to the feature size of a single channel, eliminating the bias arising from increased feature size. We then calculate the DS. 
As shown in Figure \ref{fig:ds_score_pic}, we see an increasing trend in the DS upon increasing the number of telemetry channels, indicating the increased distinguishability of malware and benign samples. This potentially indicates that the solution volume arising from $\bigcup_{i} \mathcal{V}_i$ is higher than the solution volumes of the telemetry channels $\mathcal{V}_i$ when considered independently. Next, we explore late-stage fusion-based detection to realize the potential performance improvements of incorporating multiple channels.

\subsection{Approach: Late-stage fusion}

 \begin{table}[!tb]
\centering
\caption{F1-score of the fused GLOBL channels}
\label{tab:dvfs_fusion}
\resizebox{0.3\textwidth}{!}{%
\begin{tabular}{c|cc}
\hline
\textbf{\begin{tabular}[c]{@{}c@{}}Fused channel \\ group\end{tabular}} &
  \textbf{\begin{tabular}[c]{@{}c@{}}Participating\\ Channels\end{tabular}} &
  \textbf{\begin{tabular}[c]{@{}c@{}}STD\\ Dataset\end{tabular}} \\ \hline
\textbf{DVFS} &
  1-11 &
  0.90 \\
\textbf{GLOBL} &
  1-15 &
  0.92 \\ \hline
\end{tabular}%
}
\end{table}

\begin{table}[!tb]
\centering
\caption{F1-scores of different HPC-groups before and after fusion with the DVFS and GLOBL channels}
\label{tab:hpc_dvfs_fusion}
\resizebox{0.47\textwidth}{!}{%
\begin{tabular}{cccccc}
\hline
\textbf{HPC} &
  \textbf{Standalone} &
  \textbf{\begin{tabular}[c]{@{}c@{}}With DVFS\\ (ensemble)\end{tabular}} &
  \textbf{\begin{tabular}[c]{@{}c@{}}With DVFS\\ (SG)\end{tabular}} &
  \textbf{\begin{tabular}[c]{@{}c@{}}With GLOBL\\ (ensemble)\end{tabular}} &
  \textbf{\begin{tabular}[c]{@{}c@{}}With GLOBL\\ (SG)\end{tabular}} \\ \hline
\textbf{group-1} & 0.66 & 0.89 & 0.88 & \textbf{0.93} & 0.90          \\
\textbf{group-2} & 0.71 & 0.89 & 0.84 & 0.89          & \textbf{0.92} \\
\textbf{group-3} & 0.76 & 0.90 & 0.91 & \textbf{0.93} & \textbf{0.93} \\
\textbf{group-4} & 0.67 & 0.91 & 0.90 & 0.92          & \textbf{0.93} \\ \hline
\end{tabular}%
}
\end{table}

As shown in Figure \ref{fig:feature_engineering}.(c), we consider two decision fusion approaches. In the first approach, called \textsc{ensemble}, we take a majority vote of decisions of the individual base-classifiers, each of them trained on a different telemetry channel. The second approach is a stacked generalization approach where we fuse the decisions of the base-classifiers using a second-stage model, which is a logistic regression model in our case.

\textbf{Fusing the decisions of the GLOBL channels. }
Table \ref{tab:dvfs_fusion} shows the F1-scores from fusion of different base-classifiers of the GLOBL and DVFS channels. We observe the following trends from these results: First, the F1-score obtained after fusing the GLOBL channels has a better F1-score than the sub-group of DVFS. Therefore, considering the impact of running a workload on all the sub-devices of the SoC is essential. Second, F1-score from fusion is greater than the F1-scores from each of the individual channels, which agrees with the statistical analysis in Section \ref{sssec:t_test} and validates \textit{Theorem-1}, showing fusion of distinct solution volumes of the individual telemetry channels results in a higher solution volume. 

\begin{mdframed}[style=mpdframe,frametitle=Takeaway-2]
The performance of the fused model incorporating all the GLOBL telemetry channels is greater than the performance of each channel independently, validating \textit{Theorem-1}.
\end{mdframed}

\begin{figure}[!htb]
% \vspace{-5mm}
    \centering
    \includegraphics[scale = 0.65]{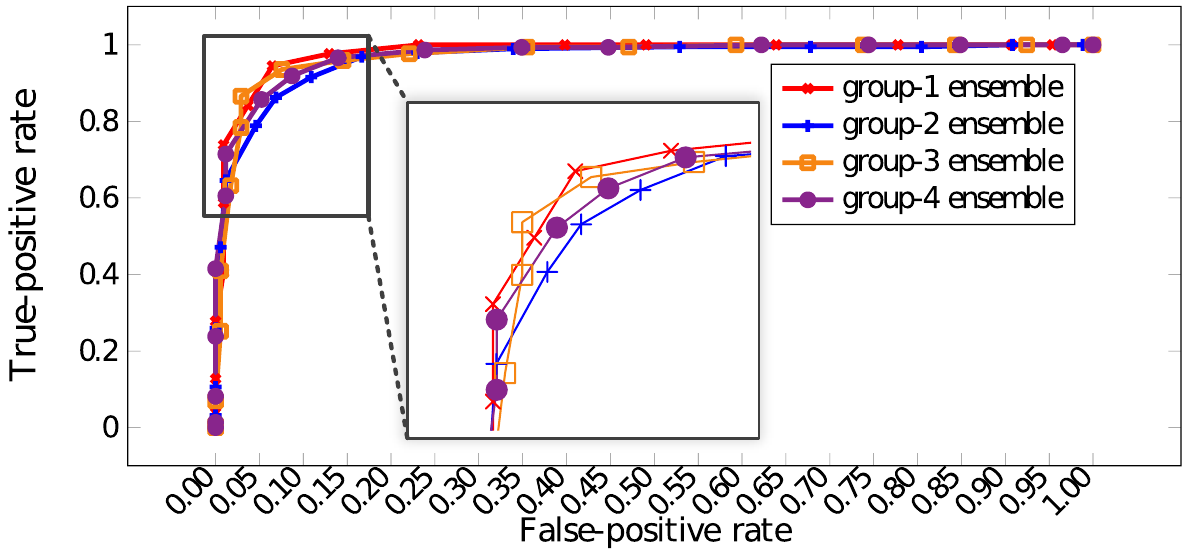}
    \caption{ROC curve for XMD}
    % \vspace{-6mm}
    \label{fig:roc}
\end{figure}

\textbf{XMD: Fusing the decisions of the GLOBL channels and HPC. }
Table \ref{tab:hpc_dvfs_fusion} shows the F1-score of the HPC groups when they were standalone and when used in conjunction with the DVFS and the GLOBL channels for both the techniques of late-stage fusion, i.e., ensemble and stacked generalization (SG). The predictive performance of the GLOBL channels, when fused with the HPCs, is higher than the fused GLOBL models (in Table \ref{tab:dvfs_fusion}) or the HPC base-classifiers. The receiver operating characteristic (ROC) curve (on the \textsc{std-dataset}) for the Ensemble-fusion models is presented in Figure \ref{fig:roc}. 

\begin{mdframed}[style=mpdframe,frametitle=Takeaway-3]
The HPC-GLOBL fused model exploits the thread-level profiling power of the HPCs and the globl profiling power of GLOBL channels, resulting in a malware detector with better predictive performance.
\end{mdframed}

\begin{figure}[!b]
% \vspace{-5mm}
    \includegraphics[scale = 0.3]{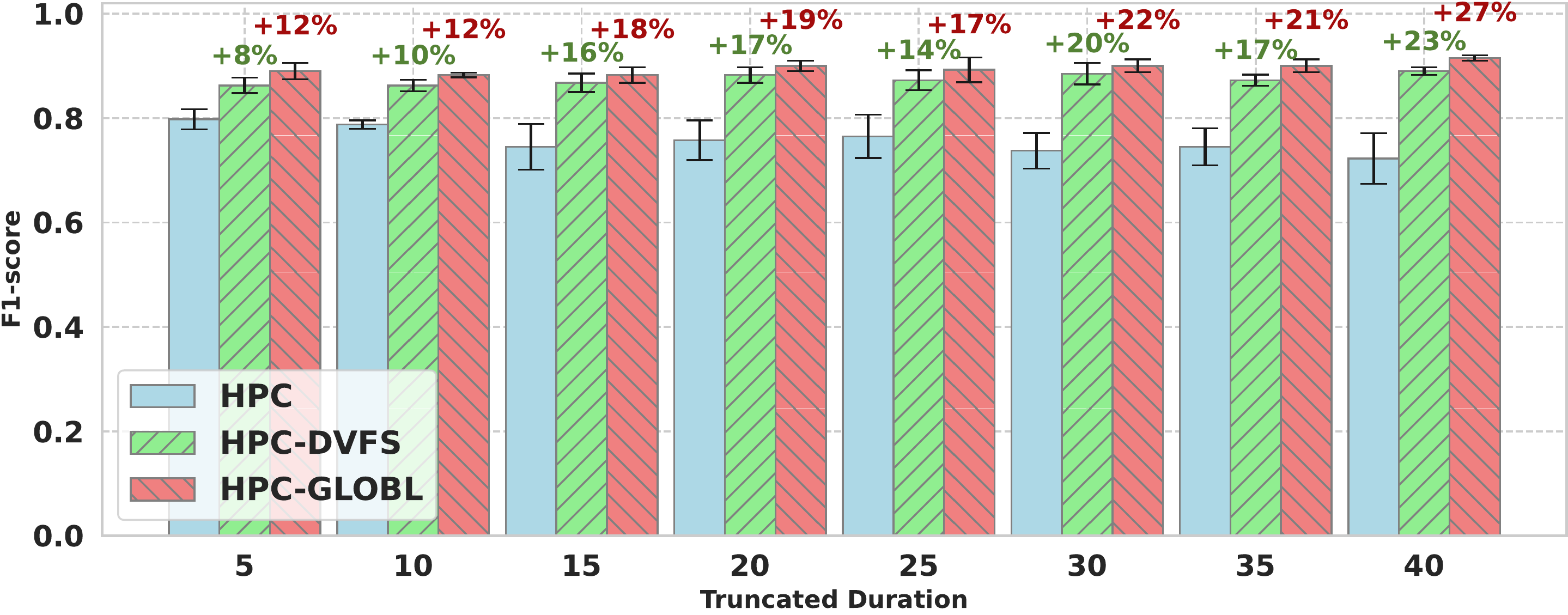}
    \caption{F1-score vs. Truncated Duration for three different HMDs: HPC, HPC-DVFS, and HPC-GLOBL.}
    % \vspace{-6mm}
    \label{fig:f1_vs_time}
\end{figure}
\subsection{Impact of execution runtime}
\label{ssec:hypothesis_vs_exec_time}
Next, we rigorously evaluate the central hypothesis of our paper by testing the impact of execution window length on the improvement in performance achieved from augmenting with the non-core telemetry. From our 40-second data-collection window, we truncated the collected data by capturing the first N seconds of hardware telemetry (referred to as 'truncated duration' from here on). We increased N in intervals of 5 seconds, and plotted the observed F1-score for three different HMDs: HPC-only, HPC-DVFS ensemble, and HPC-GLOBL ensemble, for each truncated duration in Figure \ref{fig:f1_vs_time}. The vertical spread in each bar plot accounts for consideration of all HPC-groups, with each spread corresponding to four groups. The percentage values atop the bar plots show the percentage increase in mean F1-score for HPC-DVFS and HPC-GLOBL HMDs over the HPC-only HMDs.

From this plot, two key observations can be made: (1) The HPC-DVFS and HPC-GLOBL consistently outperform the HPC-only HMDs, irrespective of the truncated duration value. This suggests that access to non-core telemetry reliably enhances the classifier's predictive performance. (2) As truncated duration increases, the percentage improvement over the HPC-only HMDs also increases. This is due to the decreasing F1-score of the HPC-only detector, countered by a slight increase for the fusion detectors, ultimately leading to a larger percentage increase. This echoes recent findings that more data does not necessarily equate to better benign vs. malware classification \cite{sandboxExecution}, a pattern we also see for the HPC-telemetry. In conclusion, the consistent percentage improvement across all truncated duration values lends strong support to our hypothesis, indicating that augmenting with non-core telemetry channels does indeed expand the attack space covered by the HMD and improve the predictive performance of the ML Classifier. Moreover, it improves the ML classifier's predictive performance across various durations of collected data.

\begin{table*}[]
\centering
\caption{Comparison of XMD against the prior works in HMD and Android-based Malware Detection}
\label{tab:comparison}
\resizebox{0.8\textwidth}{!}{%
\begin{tabular}{@{}cccc@{}}
\toprule
\textbf{Reference}                          & \textbf{Platform} & \textbf{Information}                             & \textbf{Performance} \\ \midrule
\cite{quant_hmd}          & Android           & Behavior (HPC)                                   & 0.85 TPR, 0.05 FPR   \\
\cite{demme_hpc}          & Android           & Behavior (HPC)                                   & AUC 0.82, 0.10 FPR   \\
\cite{hpc_myth_or_fact} & x86               & Behavior (HPC)                                   & 0.802 F1-score       \\
\cite{chawla_iotj}        & Android           & Behavior (CPU-DVFS)                              & 0.88 F1-score        \\ \midrule
\cite{drebin}              & Android           & Static (mainfest, disassembled code)             & 0.94 TPR, 0.01 FPR   \\
\cite{mamadroid}           & Android           & Static (API call graph)                          & 0.99 F1-score        \\
\cite{droidapiminer}       & Android           & Static (package, API information)                & 0.99 TPR, 0.022 FPR  \\
\cite{madam}               & Android           & Behavior (sys-call, critical API, user activity) & 96.9\% accuracy      \\
\textbf{Proposed} &
  \textbf{Android} &
  \textbf{Behavior (HPC, core+non-core DVFS, SYSFS)} &
  \textbf{\begin{tabular}[c]{@{}c@{}}HPC: 0.66-0.76 F1-score\\ XMD: 0.90-0.93 F1-score\end{tabular}} \\ \bottomrule
\end{tabular}%
}
\end{table*}

\begin{table*}[!htbp]
\centering
\caption{Detection rates of XMD and signature-based Antivirus Scanners on VirusTotal (\textit{\%N.P.} = Percentage of test samples on which AV scanner did not make a prediction, \textit{\%TPR} = True Positive Rate, \textit{\%FPR} = False Positive Rate)}
\label{tab:vt-comparison}
\resizebox{\textwidth}{!}{%
\begin{tabular}{cccccccccccccccccccccc}
\hline
\multirow{2}{*}{\textbf{ensemble}} & \multirow{2}{*}{\textbf{XMD}} & \multicolumn{2}{c}{\textbf{AV1}} & \multicolumn{2}{c}{\textbf{AV2}} & \multicolumn{2}{c}{\textbf{AV3}} & \multicolumn{2}{c}{\textbf{AV4}} & \multicolumn{2}{c}{\textbf{AV5}} & \multicolumn{2}{c}{\textbf{AV6}} & \multicolumn{2}{c}{\textbf{AV7}} & \multicolumn{2}{c}{\textbf{AV8}} & \multicolumn{2}{c}{\textbf{AV9}} & \multicolumn{2}{c}{\textbf{AV10}} \\ \cline{3-22} 
 &  & \textit{TPR} & \textit{\%N.P.} & \textit{TPR} & \textit{\%N.P.} & \textit{TPR} & \textit{\%N.P.} & \textit{TPR} & \textit{\%N.P.} & \textit{TPR} & \textit{\%N.P.} & \textit{TPR} & \textit{\%N.P.} & \textit{TPR} & \textit{\%N.P.} & \textit{TPR} & \textit{\%N.P.} & \textit{TPR} & \textit{\%N.P.} & \textit{TPR} & \textit{\%N.P.} \\ \hline
\multicolumn{1}{c|}{\textbf{group-1}} & \multicolumn{1}{c|}{84.13 (FPR = 3.70)} & 83.87 & \multicolumn{1}{c|}{0} & 82.76 & \multicolumn{1}{c|}{6.4} & 66.12 & \multicolumn{1}{c|}{0} & 59.64 & \multicolumn{1}{c|}{8.1} & 53.44 & \multicolumn{1}{c|}{6.4} & 53.22 & \multicolumn{1}{c|}{0} & 35.48 & \multicolumn{1}{c|}{0} & 32.25 & \multicolumn{1}{c|}{0} & 30.50 & \multicolumn{1}{c|}{4.8} & 27.42 & 0 \\
\multicolumn{1}{c|}{\textbf{group-2}} & \multicolumn{1}{c|}{87.83 (FPR = 6.89)} & 80.22 & \multicolumn{1}{c|}{3.2} & 81.93 & \multicolumn{1}{c|}{11.7} & 53.76 & \multicolumn{1}{c|}{1.1} & 60.60 & \multicolumn{1}{c|}{11.7} & 57.47 & \multicolumn{1}{c|}{7.5} & 52.17 & \multicolumn{1}{c|}{2.1} & 35.10 & \multicolumn{1}{c|}{0} & 38.46 & \multicolumn{1}{c|}{3.2} & 23.19 & \multicolumn{1}{c|}{10.6} & 23.65 & 1.1 \\
\multicolumn{1}{c|}{\textbf{group-3}} & \multicolumn{1}{c|}{86.54 (FPR = 2.90)} & 80.00 & \multicolumn{1}{c|}{0} & 79.74 & \multicolumn{1}{c|}{7.1} & 68.23 & \multicolumn{1}{c|}{0} & 48.05 & \multicolumn{1}{c|}{9.4} & 65.00 & \multicolumn{1}{c|}{5.9} & 42.86 & \multicolumn{1}{c|}{1.2} & 43.53 & \multicolumn{1}{c|}{0} & 40.00 & \multicolumn{1}{c|}{0} & 30.48 & \multicolumn{1}{c|}{3.5} & 34.11 & 0 \\
\multicolumn{1}{c|}{\textbf{group-4}} & \multicolumn{1}{c|}{85.71 (FPR = 5.23)} & 83.56 & \multicolumn{1}{c|}{0} & 76.81 & \multicolumn{1}{c|}{5.5} & 69.86 & \multicolumn{1}{c|}{0} & 54.54 & \multicolumn{1}{c|}{9.6} & 48.48 & \multicolumn{1}{c|}{9.6} & 64.28 & \multicolumn{1}{c|}{4.1} & 41.09 & \multicolumn{1}{c|}{0} & 34.78 & \multicolumn{1}{c|}{1.4} & 23.88 & \multicolumn{1}{c|}{8.2} & 29.17 & 1.4 \\ \hline
\end{tabular}%
}
\end{table*}
\textbf{Evaluating Time-to-Detection.} Detecting malware involves several stages: profiling the foreground application, performing feature engineering on the collected data, and inference using the ML classifier (malware or benign). Decisions regarding profiling duration depend on the specific use-case. As illustrated in Figure \ref{fig:f1_vs_time}, the HPC-GLOBL classifier's F1 score incrementally improves with increased execution time. Hence, the trade-off between early detection and reliable detection needs careful consideration. The next stage is feature engineering, which took 0.28s for a 40-second HPC time series and 1.12s for a GLOBL multivariate time series in our tests. The longer time for GLOBL telemetry can be attributed to the computation of the spectrogram as one of its feature engineering step. Finally, we evaluated the ML models' inference time. The random forest model used in this study exhibited a mean inference time of 0.64s over 100 iterations. The processing and inference stages thus require less time than the profiling stage. The overall time to detect malware depends on the profiling time chosen, striking a balance between detection performance and speed. This decision ultimately depends on the use-case. Furthermore, feature engineering and ML inference times can be reduced through GPU offloading, as seen in commercial HMDs \cite{intel_tdt}, or by using hardware accelerators, as prior works suggest \cite{sanyal_hmd_reservoir, analyze_hmd, map}.

\section{Comparison with Prior Works}
In this section, we compare XMD against prior works in HMD \cite{quant_hmd,demme_hpc,hpc_myth_or_fact,chawla_iotj}, software-based Android Malware Detectors \cite{drebin,mamadroid,droidapiminer,madam}, and static-analysis based Antivirus (AV) engines available on VirusTotal. These comparisons help us appreciate the efficacy of XMD in the context of existing solutions, while also guiding us in understanding its place in a system of collaborative detectors.

\textbf{Comparison with prior HMD.} We initially compare XMD with selected works in the HMD domain as shown in the first four rows of Table \ref{tab:comparison}. Notably, these works utilize HPC-telemetry as the behavioral signature, adopting a dynamic analysis-based approach \cite{quant_hmd, demme_hpc, hpc_myth_or_fact}. One also exploits CPU-DVFS as the hardware telemetry candidate \cite{chawla_iotj}. XMD outperforms these, though it's crucial to mention that the predictive performance of HPC-only and CPU-DVFS-only classifiers on our dataset (last row of Table \ref{tab:comparison}) is less than what these papers report. This discrepancy could be due to the use of benchmark applications as benigns, which may introduce bias and artificially inflate the F1-score. Additionally, our selection of malware apps influences the F1-score, as the selection process is not standardized.

\textbf{Comparison with prior software-based Android Malware Detectors.} Next, we compare XMD with specialized Android malware detectors that leverage unique information in the Android ecosystem, such as the manifest file, disassembled code, and call graphs \cite{drebin,mamadroid,droidapiminer} (next four rows in Table \ref{tab:comparison}). We also report a dynamic analysis-based approach that extracts behavioral information like sys-call, critical API, and user activity \cite{madam}. All the specialized approaches exhibit high detection performance, surpassing hardware-telemetry based approaches, including XMD. Each of these detectors, while effective, can be bypassed using specific techniques; e.g., obfuscation can bypass static analysis-based approaches. As such, a blend of detectors working on different information modalities may provide the highest security guarantees, much like the ones used in current EDR frameworks \cite{microsoft_endpoint}.

\textbf{Comparison with Signature-based AVs on VirusTotal.}
Finally, we compared XMD against the software-based AV vendors on VT that use signature-based approaches for malware detection. These included ESET-NOD32, Ikarus, K7GW, Microsoft, CAT-Quickheal, Fortinet, Avira, Cyren, Kaspersky, and Lionic (labelled as AV 1-10 in Table \ref{tab:vt-comparison}). Since VT's AV engines are signature-based, they only classify malware samples they've previously encountered. The AV engine does not make a prediction on the samples it has not been trained on (called zero-day malware). We only considered vendors that provided decisions for at least 85\% of the samples in our test dataset.  As shown in Table \ref{tab:vt-comparison}, while the best detection performance among the AVs on the test data for group-3 is around 80\%, XMD achieves a detection rate of 86.54\% with a false positive rate of 2.9\%. The table also includes a \%N.P. value for each AV, indicating the percentage of test samples (zero-day) the AV did not make a prediction for.

\textbf{Limitations of comparison against prior works.} A direct comparison with any prior dynamic analysis-based approach is challenging due to the costly execution of samples on the bare-metal framework, non-disclosure of malware and benign samples' hashes in prior works, and the incompatibility of older, released samples with the newer OS in our framework. Therefore, we cannot provide a direct comparison and just report the numbers that were reported in the respective works.

\begin{mdframed}[style=mpdframe,frametitle=Takeaway-4]
XMD outperforms previous CPU-telemetry-based HMDs and signature-base AVs on VirusTotal; it trails specialized Android-based detectors.
\end{mdframed}

\section{Discussion, Limitations, and Future Work}
\label{sec:discussion}

\textbf{Non-determinism in Hardware Telemetry. } Recent works have identified failure scenarios in HMDs due to the non-determinism in the HPC-based telemetry arising from interrupt-skid and overcounting of instructions \cite{sok_hpc_malware_detect, hpc_myth_or_fact}. These proof-of-concept attacks are aimed at skewing the CPU telemetry measurements. Since XMD relies on a diverset set of core and un-core telemetry channels, it is potentially robust to such proof-of-concept attacks that tamper with the CPU telemetry measurements. Future works will explore the XMD's resilience against collusion-based attacks that tamper with the hardware telemetry. 

\textbf{Non-Core Telemetry Limitations and Extension to Desktop Devices.} The utilization of system-wide telemetry channels to monitor non-core devices poses a risk, as it leaves the defense susceptible to user-land bad actors. These actors can craft collusion-based attacks, injecting noise into the telemetry channels. A potential remedy is the adoption of non-core performance counters that offer process-level tracking of non-core devices. These counters, currently employed by hardware vendor profiling frameworks, monitor diverse activities such as traffic to storage and network devices, memory access patterns like DRAM and cache load stalls, GPU and CPU usage in profiled applications \cite{intel}. Repurposing these tools' drivers can enhance deployed HMDs' detection performance \cite{intel_tdt}. While the HPC-events for CPUs are generally standardized across platforms, the availability of uncore-counters is inconsistent, absent on client-side PCs but present on select server-class platforms. Hence, there's a call for an industry-wide initiative to integrate uncore telemetry into existing HMD frameworks. As our study concentrates on mobile devices, extending our approach to desktop devices requires a comparable process-level tracking solution for non-core telemetry. This is due to the potential masking of individual process behaviors by concurrent foreground applications in desktop environments when using system-level profiling. Thus, the extension of XMD to desktop devices is reserved for future development and exploration.

\textbf{Data-Collection Environment Limitations.} Despite our efforts to make our data-collection environment realistic for mobile devices, there are corner cases that we haven't addressed. We employ the Monkey tool for Android application interactions, which uses stateless interactions, differing from real user inputs. An improved approach for simulating user interface interactions is to use a state-based model like Droidbot \cite{droidbot}. However, in our setup, Droidbot exhibited unpredictable crashes, necessitating human intervention. Future work could explore better methods for simulating human interactions for extensive telemetry collection. Our study's analysis is confined to a single Google Pixel 3 mobile device, limiting the scope regarding SoC chipsets and mobile devices. Nevertheless, the theorems and empirical observations in our work are not dependent on a specific architecture or platform and could potentially extend to other mobile devices. We plan to empirically verify our proposed theorems on different mobile device platforms in future work. While we have outlined our environment collection assumptions—involving a user interacting with one foreground application while multiple system-level workloads run in the background (in Section \ref{ssec:environment_description})—a more rigorous set of environment assumptions can be tested to robustly evaluate our proposed hypothesis. These could include distinct user profiles, scenarios with multiple apps, background apps, or extended execution time. In the scope of this study, a comprehensive long-term evaluation was not conducted, a limitation we acknowledge. The exploration of these aspects, including the potential impact of a long-term evaluation on the system's performance, is reserved for future investigations.

\textbf{Dataset Size: A Comparison with Software-based Approaches.} We must highlight the relatively small size of our dataset, especially when compared to static analysis-based datasets like MAMADroid \cite{mamadroid} and DREBIN \cite{drebin}. Static analysis techniques offer notable scalability as they do not require malware execution. Nevertheless, their susceptibility to evasion via simple obfuscation is a significant disadvantage. In comparison with other software-based, dynamic-analysis methods, our dataset remains modest in size \cite{madam}. Techniques like syscall-based malware detectors, used in behavioral analysis, employ virtualized sandboxes to improve scalability in data collection. However, HMDs require data collection on bare-metal devices, since virtual machines potentially introduce errors \cite{hpc_myth_or_fact}, and there are limited HPC events that can be accessed from within a virtual machine. As a result, scalabiltiy of data-collection is a major limitation in the design of any HMD. Thus, our dataset size aligns with those considered in previous HMD studies \cite{hpc_myth_or_fact, quant_hmd, demme_hpc}.

\textbf{Choice of fusing information from multiple telemetry channels.} We have used a late-stage fusion approach to incorporate the power of multiple telemetry channels. However, an intermediate fusion-based approach can potentially result in better detection performance at the cost of increased complexity, e.g., an approach that exploits the interaction between the different telemetry channels. We leave the exploration of such novel ML agents as future work.

\textbf{Future Work.} Future work will extend across several dimensions. We aim to probe XMD's resilience against more complex, collusion-based attacks that interfere with hardware telemetry. Furthermore, the development and exploration of XMD for desktop devices will be undertaken, requiring the devising of a process-level tracking solution for non-core telemetry. We also plan to improve our data collection environment, by seeking better methods to simulate human interactions and by testing our hypotheses across various mobile device platforms and under a more comprehensive set of environment assumptions. Lastly, we plan to investigate the potential of novel machine learning agents capable of exploiting interactions between different telemetry channels for more efficient malware detection.

\section{Conclusion}
In this paper, we propose XMD, a hardware-based malware detector that uses the expansive set of hardware telemetries from the different sub-systems of an SoC used in a mobile device. We develop a theorem that leverages the Replica Theory of object manifolds to establish the performance gains expected from incorporating the expansive set of telemetry channels. We evaluate XMD using a dataset containing the expansive hardware telemetry of malware and benign applications collected from a commodity smartphone. Our findings suggest that XMD outperforms the current CPU-telemetry-based malware detectors (HPC-based and CPU-DVFS-based). We also show that XMD provides better detection performance than the commercial AV software on VirusTotal with acceptable false-positive rates; however XMD trails in performance when compared to specialized Android malware detectors. Since XMD relies on a different modality of information when compared to software-based detectors, XMD can complement other software-based detection approaches in a collaborative defense system.

\bibliographystyle{IEEEtran}
{\footnotesize
\bibliography{ref.bib}}

\end{document}